\documentclass[%
 reprint,
showkeys,
 amsmath,amssymb,
 aps,
]{revtex4-1}

\usepackage{graphicx}
\usepackage{dcolumn}
\usepackage{bm}

\renewcommand{\vec}[1]{\mbox{\boldmath $#1$}}
\renewcommand{\tensor}[1]{\mbox{\boldmath $#1$}}

\newcommand{\tr}{\mathrm{tr}}

\begin{document}


\title{
Strain Mode of General Flow:
\\
Characterization and Implications for Flow Pattern Structures
}%

\author{Yasuya Nakayama\(^{1}\)}
\email{nakayama@chem-eng.kyushu-u.ac.jp}
\author{Tatsunori Masaki\(^{2}\)}
\author{Toshihisa Kajiwara\(^{1}\)}
\affiliation{%
\(^{1}\)Department of Chemical Engineering,
Kyushu University,
Nishi-ku,
Fukuoka 819-0395,
Japan
}%
\affiliation{%
\(^{2}\)Development and Production Department, UNITIKA Ltd., 31-3, Uji-Hinojiri, Uji-shi, Kyoto 611-0021, Japan
}


\date{\today}

\begin{abstract}
Understanding the mixing capability of mixing devices based on their geometric
shape is an important issue both for predicting mixing
processes and for designing new mixers.
The flow patterns in mixers are directly connected with the modes of the local
strain rate, which is generally a combination of elongational flow and
planar shear flow.
We develop a measure to characterize the modes of the strain rate for
general flow occurring in mixers.
The spatial distribution of the volumetric strain rate (or non-planar
strain rate) in connection with the flow pattern 
plays an essential role in understanding distributive mixing.
With our measure, flows with different types of screw elements in a
twin-screw extruder are numerically analyzed.
The difference in flow pattern structure between conveying screws and kneading
disks is successfully characterized by the distribution of the volumetric
strain rate.
The results suggest that the distribution of the strain rate mode offers
an essential and convenient way for characterization of the relation
 between flow pattern structure and the mixer geometry.
%
\end{abstract}

\maketitle


%
\section{Introduction}
Mixing is one of the most important process in industries, including
polymer processing, rubber compounding, and food processing, because it
directly affects the quality of multi-component
materials~\cite{Tadmor2006Principles,White2010Twin,2011Food,Rauwendaal2014Polymer}.
Several types of mixing devices, such as twin-screw extruders, twin-rotor
mixers, and single-screw extruders, have been developed for different material
processabilities and different processing purposes.
To select an appropriate mixing element for a variety of complex
materials and product qualities, one fundamental issue is the fluid
mechanical characterization of the mixing capability of various types of
the mixing elements.

The mixing process, or, more specifically, the reduction of the inhomogeneity of
material mixtures, is achieved through material flow driven by mixing
devices. Therefore, in principle, the capability of the mixing elements can
be evaluated through an analysis of the flow in the device.
Along these lines, the visualization of the flow patterns has been performed
numerically or experimentally to obtain a qualitative insight into
the global mixing
kinetics~\cite{Lawal1995Mechanisms,Carneiro1999Flow,Funatsu20023D,Ishikawa2002Flow,Zhang2009Numerical,Kubik2012Method,SarhangiFard2013Simulation}.
While the global flow pattern characterizes the evolution of the material
distribution, the divergence of the material trajectories is locally
associated with the deformation of the fluid elements.
Since the substantial local deformation rate is described by the strain rate
tensor, several approaches for the characterization of the local strain rate
have been
developed~\cite{Ottino1979Lamellar,Ottino1989Kinematics,ManasZloczower1996Analysis,Nakayama2011Meltmixing}.

The degree of irrotationality of the deformation rate, which is often
called the ``mixing index''~\cite{Connelly2007Examination}, has been used to
quantify dispersive mixing
efficiency~\cite{Ottino1989Kinematics,ManasZloczower1996Analysis}.  This characterization was
motivated by the experimental fact that elongational flows are more
effective than simple shear for droplet/agglomerate breakup.  In
a rheometric flow setup, the elongational flows are irrotational, while the
simple shear flow is a superposition of a planar shear flow and a
rotational flow in a plane.  Therefore, the irrotationality can define
the flow type for the rheometric flow setup.
However, in generic flows in mixing devices, the type of elongational
flows described by the strain rate tensor cannot be assessed by the
irrotationality by definition.
Another quantity, called the mixing efficiency, is the relative magnitude
of the elongational rate along a certain direction to the magnitude of
the strain rate tensor~\cite{Ottino1979Lamellar,Ottino1989Kinematics}.
The mixing efficiency is useful when the interface between phases is
well defined.
If, in the mixing efficiency, the maximal elongational direction is taken,
the modes of the strain rate, such as the uniaxial/biaxial elongational
flows and the planar shear flow, are
discriminated~\cite{Nakayama2011Meltmixing}.  In this case, the eigenvalue
problem for the strain rate tensor should be solved.
The distribution of the strain-rate modes in combination of the
magnitude of the strain-rate in general flows
should be useful in understanding the dispersive mixing
capability. 
The strain-rate modes in three-dimensional flows can be
identified in principle by combining the different eigenvalues of the
strain-rate tensor, and thus different forms for quantification of the
strain-rate mode can be designed. However, such quantification has not
been derived in a numerically tractable manner.

To the best of our knowledge, the role of elongational flow in
distributive mixing has been only rarely discussed, whereas the
effectiveness of irrotational elongational flows for dispersive mixing
has been well recognized.
Flow trajectories diverge by elongational flows 
however small the elongational rate is.  
The divergence
directions are restricted to a plane for planar shear flow, but involve
three directions for non-planar elongational flows.
If distributive mixing is effectively promoted in the confined space of
a mixing device during a finite period of time, 
a flow pattern being effective to distributive mixing should be
developed.
Such a flow pattern is expected to be associated with a
certain distribution of the non-planar or volumetric elongational flows.
The distribution of the volumetric flows is therefore
considered to give essential information for a better understanding of
the mixing capability of mixing elements.

Especially, the flow patterns in the regions of smaller strain rates 
largely determine net mixing capabilities of mixing devices. 
Although the area-stretching ability is very low in the small strain
rate regions, such regions occupy a large fraction of the channel, and
the materials are conveyed to larger strain rate regions by the flow in
the small strain rate regions.
The distribution of the volumetric flows is expected to be useful in
understanding the relation between the structure of the flow patterns in
the small strain regions and the mixer geometry.

In general, the flow field in a mixing device is an arbitrary combination of
volumetric elongational flows and a planar shear flow.
Thus, for a general three-dimensional flow, rather than only rheometric flows,
the characterization of volumetric elongational flows from the strain rate
tensor is required.
Although the strain-rate mode should be an important quantity both in
dispersion capability and analysis of flow-pattern structures in
relation to distributive mixing, 
its characterization in three-dimensional flows has not established so far.

In this paper, we derive a measure which identifies the volumetric
elongational flows from the strain rate tensor.
We apply this measure to melt-mixing flow in twin-screw extrusion.
Using the distribution of the volumetric elongational flows, we discuss
the differences in the flow patterns and the mixing characteristics of
the different screw elements.

\section{Theory}
In any kind of flow in a mixing device, the local deformation rate is a
combination of shear flow, elongational flow, and rotational flow.
The deformation rate of a fluid element, \(\tensor{\nabla}\vec{v}\), is
decomposed into the strain rate tensor \(\tensor{D}\) and the vorticity
tensor \(\tensor{\Omega}\):
\begin{align}
 \vec{\nabla}\vec{v}
&= \tensor{D}+\tensor{\Omega},
\end{align}
where 
\(\vec{v}\) is the velocity field,
\(\tensor{D}=
\left(
\vec{\nabla}\vec{v}
+
\vec{\nabla}\vec{v}^{T}
\right)/2
\),
and
\(\tensor{\Omega}=
\left(
\vec{\nabla}\vec{v}
-
\vec{\nabla}\vec{v}^{T}
\right)/2
\),
the superscript \(T\) indicates the transpose.
Concerning mixing processes, the change in the distance between two nearby points
is represented by \(\tensor{D}\).  Thus, the strain rate tensor is
mainly responsible for the local mixing capability.

For incompressible flows, {\itshape i.e.}, \(\tr\tensor{D}=0\), the
strain rate tensor can be diagonalized with an orthonormal matrix
\(\tensor{P}\),
\begin{align}
 \tensor{D}
&=
 \tensor{P}^{T}
\cdot
\tensor{\Lambda}
\cdot
 \tensor{P},
\\
\tensor{\Lambda}
&=
\begin{pmatrix}
 \dot{\varepsilon} & 0 & 0\\
0 &  (m-1)\dot{\varepsilon}&  0\\
 0 & 0 & 
-m\dot{\varepsilon}
\end{pmatrix}
,
\end{align}
where \(\dot{\varepsilon}>0\) and \(-m\dot{\varepsilon}\), respectively,
represent the largest and the smallest eigenvalues of \(\tensor{D}\) by
assuming \(1/2\leq m\leq 2\).
The value of \(m\) specifies the mode of the strain rate. For example,
\(m=1/2\) for a uniaxial elongational flow, \(m=2\) for a biaxial
elongational flow, and \(m=1\) for a planar shear flow.

A typical value of the strain rate is \(\dot{\varepsilon}\), but the magnitude
of the strain rate is commonly evaluated by the second invariant of
\(\tensor{D}\) without explicit calculation of the eigenvalues,
\begin{align}
\label{eq:dd}
\tensor{D}: \tensor{D}
&=
2(1-m+m^{2})\dot{\varepsilon}^{2},
\end{align}
because it is always positive for finite values of \(\dot{\varepsilon}\)
by definition.
Since we are interested in the mode of the strain rate, or, equivalently,
the value of \(m\) without solving the eigenvalue problem, we consider
another invariant of \(\tensor{D}\).
The determinant of \(\tensor{D}\), a third-order invariant, is expressed
in terms of the eigenvalues by
\begin{align}
\label{eq:determinant}
 \det\tensor{D}
&=
m(1-m)\dot{\varepsilon}^{3}.
\end{align}
From Eq.~(\ref{eq:determinant}), we can see that the determinant of
\(\tensor{D}\) becomes zero for planar shear (\(m=1\)), irrespective of
\(\dot{\varepsilon}\).
In other words, \(\det\tensor{D}=0\) means that the strain directions
are restricted to be within a certain plane, and \(\det\tensor{D}\neq 0\) indicates
that the strain directions extend three-dimensionally.  With this
property of \(\det\tensor{D}\), we can get an insight into whether the
local strain rate is more planar or more volumetric without explicitly
calculating the eigenvalues of \(\tensor{D}\).

Combining Eqs.~(\ref{eq:dd}) and (\ref{eq:determinant}), we define a
measure for the mode of the strain rate, independent of the value of
\(\dot{\varepsilon}\), by
\begin{align}
\beta &=
\frac{3\sqrt{6}\det\tensor{D}}{
\left(
\tensor{D}:\tensor{D}
\right)^{3/2}}\,,
\end{align}
where the numerical prefactor is chosen so that the range of \(\beta\)
is normalized to \([-1,1]\).  Special cases include: \(\beta=-1\) for a
biaxial elongational flow, \(\beta=1\) for a uniaxial elongational
flow, and \(\beta=0\) for a planar shear flow.  For the general case of
\(0<\left|\beta\right|<1\), the strain rate is an arbitrary
superposition of an elongational flow and a shear flow.  Both
\(\tensor{D}:\tensor{D}\) and \(\det\tensor{D}\) are frame invariant,
implying that \(\beta\) is frame invariant as well, namely \(\beta\) does
not change in different frames of the observers.

Now we consider a relation between the strain rate mode and the mixing capability.
In general, if a mixing device has good mixing capability, the flow in
the device should satisfy the following two conditions.  
(i) There exist
some regions with large strain rates in order for the materials to be
stretched extensively.  (ii) All the fluid elements in the mixer are
efficiently and repeatedly
conveyed to the large strain rate regions in order to have chances to be
stretched. 
The distribution of the
magnitude of the strain rate of Eq.~(\ref{eq:dd}) (or in the form of
shear stress,
\(\eta(\tensor{D}:\tensor{D})\sqrt{\tensor{D}:\tensor{D}}\), with the
shear viscosity \(\eta\) as a function of the magnitude of the
strain rate) has often been discussed
\cite{Ishikawa2002Flow,Zhang2009Numerical,Yao1998Influence,Alsteens2004Parametric,Bravo2004Study,Malik20053D,Connelly20063D,Rathod2013Effect}.
In mixing devices for highly viscous fluids, like polymers, 
the largest strain rate is achieved in narrow gap regions, which are the
clearance between the screw/rotor tips and the barrel surface, and the gap
between two screws/rotors in twin-shaft machines.
Thus, 
it is usually the case the condition (i) is satisfied. However, it is
only a necessary condition for good net mixing capability.
These large strain rate regions are also responsible for the dispersion of
agglomerates.

In contrast to the large strain rate regions, 
 a role of regions with smaller strain rates on net
distributive mixing 
is rather
difficult to characterize because this would, in principle, require an
analysis of the ensemble of trajectories.
Nevertheless, the small strain rate regions are mainly responsible for
the condition (ii).
In the past, distributive mixing was discussed using tracer tracking and
Lagrangian
statistics~\cite{Lawal1995Mechanisms,Funatsu20023D,Ishikawa2002Flow,Zhang2009Numerical,Nakayama2011Meltmixing,Connelly2007Examination,Bravo2004Study,Lawal1995Simulation,Yao1996Analytical,Yao1997Design,Cheng1998Distributive,Yao2001Mixing,Hirata2014Effectiveness}.
These methods are basically for the characterization of the global properties of
a mixing device, and are not for the direct characterization of a
relation between the local flow characteristics induced by a screw geometry
and the mixing capability.
Effective and repetitive transport of the materials to the large strain
rate regions is one important role of the small strain rate region, and 
it affects net mixing performance of the mixers.
This flow pattern structure in the small strain rate regions is
basically based on the mixer geometry.

In order for the fluid elements to be stretched and folded over the whole
space in a mixing device, the flow trajectories should be bifurcated and
change directions with some frequency during the mixing process.
This bifurcation of the flow trajectories is directly related to the
volumetric modes of the strain rate.  Therefore, the spatial
distribution of the strain rate mode gives directly an insight into the
relation between the mixing process and the flow driven by the mixer
geometry.

For the characterization of the local deformation rate, there is another quantity,
called the mixing index~\cite{Cheng1990Flow},
\begin{align}
\label{eq:mz}
 \lambda &= \frac{\sqrt{\tensor{D}:\tensor{D}}}{
\sqrt{\tensor{D}:\tensor{D}}
+
\sqrt{\tensor{\Omega}:\tensor{\Omega}^{T}}
}\,,
\end{align}
which has often been used to characterize the efficiency of the dispersive mixing
of mixing
devices~\cite{Yao1998Influence,Connelly20063D,Rathod2013Effect,Yang1992Flow,Cheng1997Study}.
Some authors have used different normalization like \(
\left(
\sqrt{\tensor{D}:\tensor{D}}-\sqrt{\tensor{\Omega}:\tensor{\Omega}^{T}}
\right)
/
\left(
\sqrt{\tensor{D}:\tensor{D}}+\sqrt{\tensor{\Omega}:\tensor{\Omega}^{T}}
\right) 
\) which takes a value in
\([-1,1]\)~\cite{Ottino1989Kinematics,Jongen2000Characterization} while
\(\lambda\) is normalized to be within \([0,1]\).
Since \(\tensor{\Omega}:\tensor{\Omega}^{T}\) is not frame invariant
when the observer's frame depends on time, \(\lambda\) is not frame
invariant in general. Nevertheless, if we choose a time independent
frame like the laboratory system of coordinates, \(\lambda\) is
practically regarded as objective.  By the definition in Eq.~(\ref{eq:mz}),
\(\lambda\) defines the degree of irrotationality of the deformation rate,
\textit{e.g.}, \(\lambda=0\) for pure rotation, \(\lambda=1\) for an irrotational
flow, and \(0<\lambda<1\) for a partially rotational flow; furthermore, \(\lambda\)
is independent of the strain rate mode.  
In two-dimensional flows, since available strain-rate mode is only planar
elongational flow, irrotationality can be used to distinguish between
simple shear flow (\(\lambda=1/2\)) and elongational flow (\(\lambda=1\)).
However, this is not the case for three-dimensional flows since different
strain-rate modes exist. 
In three-dimensional flows, \(\lambda\) simply indicates the
irrotationality, but not the strain-rate mode.
The mixing index \(\lambda\)
and our measure for the strain rate mode \(\beta\)
characterize different aspects of the local deformation rate.

\section{Numerical Simulation}
The flow of a polymer melt in the melt-mixing zone of a twin-screw extruder
has been numerically solved to allow understanding the relation between
the distribution of the strain rate mode and the mixing capability of screw
elements.
The configuration of the screws is shown in Fig.~\ref{fig1}.
%
From the inlet, a forwarding conveying screw, kneading disks, and
a backwarding conveying screw are arranged.
The melt-mixing zone consists of kneading disks of neutrally staggered
five blocks~\cite{Kohlgruber2007CoRotating}.
The diameter of the barrel is set to
\(D=69\)\;mm, the length of the computational domain is \(L=2.92 D\).

We focus on the situation where the material fully fills the channel.
The Reynolds number is assumed to be much less than unity, so that
inertial effects are neglected.
The flow is assumed to be incompressible, and in a pseudo-steady state to
screw rotation, as has often been assumed in polymer flow in twin-screw
extruders~\cite{Ishikawa20003D,Bravo2004Study,Malik20053D,Zhang2009Numerical,Nakayama2011Meltmixing,SarhangiFard2013Simulation,Rathod2013Effect,Hirata2014Effectiveness}.
With these assumptions, the governing equations become
\begin{align}
 \vec{\nabla}\cdot\vec{v} &= 0, 
\\
0 &= -\vec{\nabla}p+ \vec{\nabla}\cdot\tensor{\tau},
\\
\rho c_{p}\vec{v}\cdot\vec{\nabla}T &= k\nabla^{2}T + \tensor{\tau}:\tensor{D},
\end{align}
where \(p\) is the pressure, \(\tensor{\tau}\) is the deviatoric stress,
\(\rho\) is the mass density, \(c_{p}\) is the specific heat capacity,
\(T\) is the temperature, and \(k\) is the thermal conductivity.

The fluid is assumed to be a viscous shear-thinning fluid that follows
Cross--Arrhenius viscosity~\cite{Cross1965Rheology}, 
\begin{align}
 \tensor{\tau} &= 2\eta\tensor{D},
\\
\eta &= \frac{\eta_{0}}{1+c\left(\eta_{0}\dot{\gamma}\right)^{n}},
\\
\eta_{0} &= a\exp\left(\frac{b}{T}\right),
\\
\dot{\gamma} &= \sqrt{2\tensor{D}:\tensor{D}},
\end{align}
whose parameters are
obtained by fitting the shear viscosity of a polypropylene melt taken
from~\cite{Ishikawa20003D}, and the values are \(c=1.3575\times
10^{-3}\), \(n=0.66921\), \(a=1.7394\), and \(b=4656.8\).
The mass density, specific heat capacity, and thermal conductivity are
taken from~\cite{Ishikawa20003D} as well, and the values are
\(\rho=735.0\)\;kg/m\(^{3}\), \(c_{p}=2100\)\;J/(kg\(\cdot\)K), and
\(k=0.15\)\;W/(m\(\cdot\)K).  

As operational conditions, the volume flow rate and the screw rotation
speed are set to 60\;cm\(^{3}\)/s (\(\approx 159\)\;kg/h)
 and 200\;rpm.  
The no-slip condition on the velocity at the barrel and screw
surfaces is assumed. The velocity at the inlet boundary
was set to a value under the given volume flow rate.
The pressure at the outlet boundary was fixed to be a constant value.
The temperatures on the barrel surface and at the
inlet boundary were set to be a constant value of 473.15\;K. The natural
boundary conditions for the temperature equation in the exit
boundary plane and the screw surface were assigned.

The set of equations was discretized by the finite volume
method and solved by SIMPLE method~\cite{Versteeg2007Introduction}
using a commercial software, ``R-FLOW'' (R-flow Co., Ltd., Saitama,
Japan).
From the obtained velocity field, the strain rate and the
vorticity tensors were calculated so as to evaluate \(\beta\) and
\(\lambda\).

\newcommand{\figone}{Screw configuration of melt-mixing zone.  From the
inlet (bottom of the figure), forwarding conveying screw~(FS),
neutrally-staggered kneading disks~(N-KD), and backwarding conveying
screw~(BS). The extrusion direction is indicated by the arrow on the
right, while the rotation direction is indicated by the arrows on the
top.}

\begin{figure}
 \center \includegraphics[width=.5\hsize]{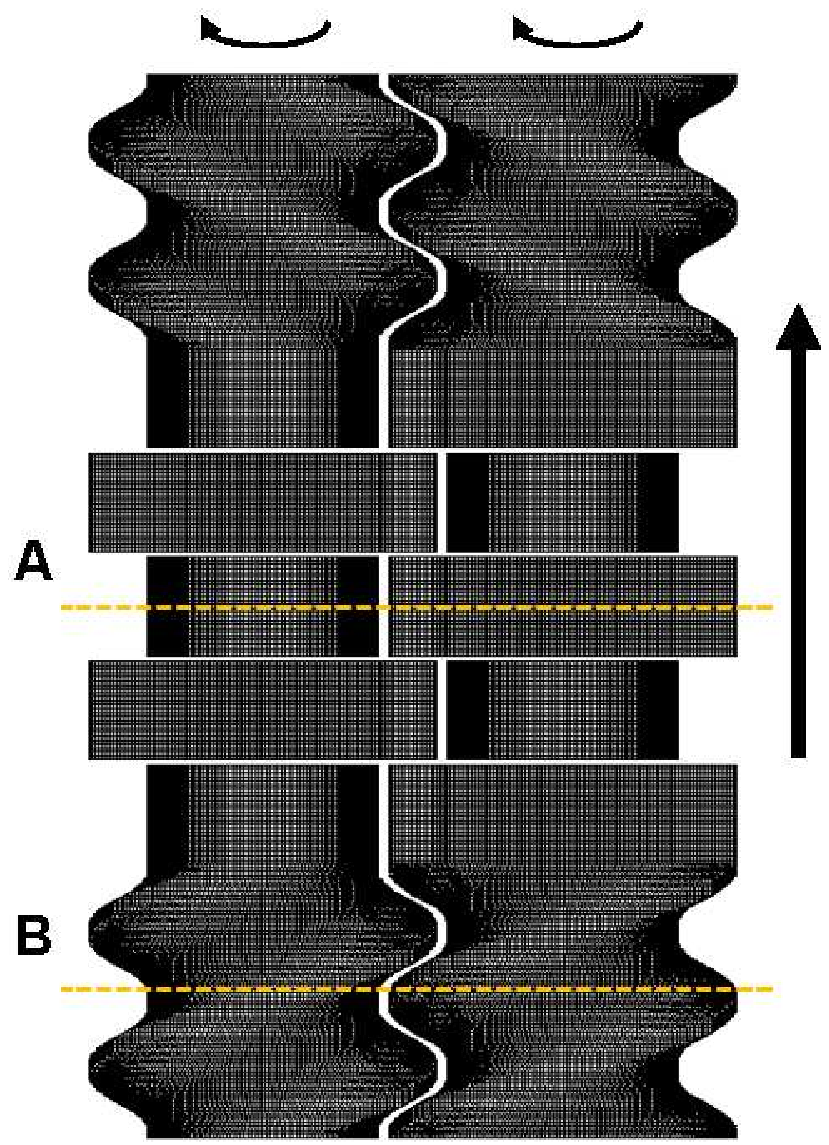}
\caption{\texttt{\figone}
}
\label{fig1}
\end{figure}

\section{Results and Discussion}
For typical twin-screw mixers, the largest strain rate occurs mainly around
the tips and the gap between the two screws in the intermeshing region
because the channel widths of these regions are designed to be the smallest
in the device.
This is common for the conveying screw of extruders and
many types of kneading elements, such as kneading disks, rotor elements,
screw mixing elements, and so on~\cite{Lawal1995Mechanisms,Carneiro1999Flow,Funatsu20023D,Ishikawa2002Flow,Zhang2009Numerical,Kubik2012Method,SarhangiFard2013Simulation,Nakayama2011Meltmixing,Yao1998Influence,Alsteens2004Parametric,Bravo2004Study,Malik20053D,Rathod2013Effect,Hirata2014Effectiveness,Ishikawa20003D,Kalyon2007Integrated,Vyakaranam2012Prediction}.
To observe this, 
Fig.~\ref{fig2} shows the distribution of the magnitude of
the strain rate, \(\sqrt{\tensor{D}:\tensor{D}}\), at the cross section B
indicated in Fig.~\ref{fig1}.
At the clearances between the screw-tips and the barrel surface, and the gap between
two screws, \(\sqrt{\tensor{D}:\tensor{D}}\) takes values larger than
600\;\((1\)/s), which is consistent with the circumferential shear rate of 460\;\((1\)/s)
estimated from the screw rotation speed of 200\;rpm.
In contrast, the magnitude of the strain rate in other regions is
extremely small.  The distribution of
\(\sqrt{\tensor{D}:\tensor{D}}\) is basically similar for the conveying screws and
the kneading disks shown in Fig.~\ref{fig1}, suggesting that the
difference in the mixing capabilities of these elements are largely
attributable to the differences in the flow patterns in the small strain rate
regions.

\newcommand{\figtwo}{Distribution of
\protect\(\protect\sqrt{\protect\tensor{D}:\protect\tensor{D}}\protect\)
as a typical magnitude of the strain rate at position B in
Fig.~\ref{fig1}.
Large values of
\protect\(\protect\sqrt{\protect\tensor{D}:\protect\tensor{D}}\protect\)
over 600\;(1/s) are reached at the narrow gaps between the screw tips
and the barrel surface and between the two screws, while in the other
regions, the value of
\protect\(\protect\sqrt{\protect\tensor{D}:\protect\tensor{D}}\protect\)
stays rather small.  }

\begin{figure}
 \center
\includegraphics[width=\hsize]{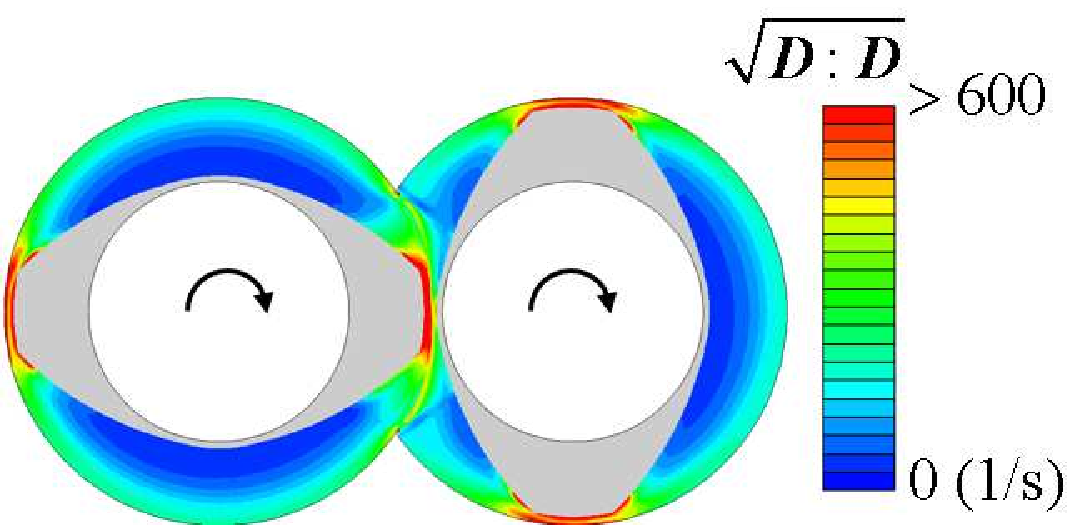}
\caption{\texttt{\figtwo}
}
\label{fig2}
\end{figure}

Next, we compare the distributions of the volumetric strain rate between the
conveying screw and the kneading disks.
We use the absolute value of \(\beta\) as defining 
\(d_{v}=\left|\beta\right|\), which is zero for a planar shear, 
unity for pure elongational flows, and becomes
\(0<d_{v}<1\) for an arbitrary superposition of planar and
elongational flows,  for we are interested in how the volumetric
strain rate is distributed.
Figure~\ref{fig3} shows the distributions of \(d_{v}\)
at a cross section in the conveying screw (location B in
Fig.~\ref{fig1}) and at a cross section in the kneading disks (location A in
Fig.~\ref{fig1}).
In the case of the conveying screw shown in Fig.~\ref{fig3}(a),
although an elongation flow occurs in some small regions around the
tip-barrel clearance and in the intermeshing regions,  small values of
\(d_{v}\) prevail in most of the section.
This was observed in the other phase of the screw rotation.
From this observation, circumferential planar shear is predominant in the flow driven by
a conveying screw, which  corresponds to the flow along the screw root.
The elongational flows around the tip-clearance regions in the conveying screw are explained as follows.
When fluid elements go across a screw tip, that flow is a bifurcation from
the upper stream flow along
the screw root, followed by the confluence to another flow along the
next screw root. These flow patterns are observed as the elongational
flows in Fig.~\ref{fig3}(a). Except for these flow patterns, interchange of
materials rarely occurs along screw rotations.
This fact is consistent with the well-known low level of the mixing
capability of a conveying screw, because a volumetric bifurcation of
the trajectories rarely occurs, so that the distributive mixing is not
much promoted.

In the case of the kneading disks shown in Fig.~\ref{fig3}(b),
we found that the volumetric strain rate develops in a remarkably large fraction of
the section, and its distribution forms a characteristic pattern. 
Elongational flows occur in the region far from the
screw tips and the surfaces, which are, coincidentally, small strain rate regions.
The locations of the elongational flows correspond to those of
the opening space between neighboring staggered disks.
As demonstrated in Fig.~\ref{fig3}, the distribution of the volumetric
strain rate closely reflects the flow pattern structure caused by the
geometric shapes of the elements, and
characterizes the flow pattern in the small strain rate regions.

\newcommand{\figthree}{Distributions of the degree of volumetric strain
rate
\protect\(d_{v}=\protect\left|\protect\beta\protect\right|\protect\) in
the melt-mixing zone.  The value of \protect\(d_{v}\protect\) is zero
for a planar shear, unity for a pure elongational flow, and
\protect\(0<d_{v}<1\protect\) for a general flow.  (a) The section in a
conveying screw at position B in Fig.~\ref{fig1}, (b) The section
in the kneading disks at position A in Fig.~\ref{fig1}.  }

\begin{figure}
 \center
\includegraphics[width=\hsize]{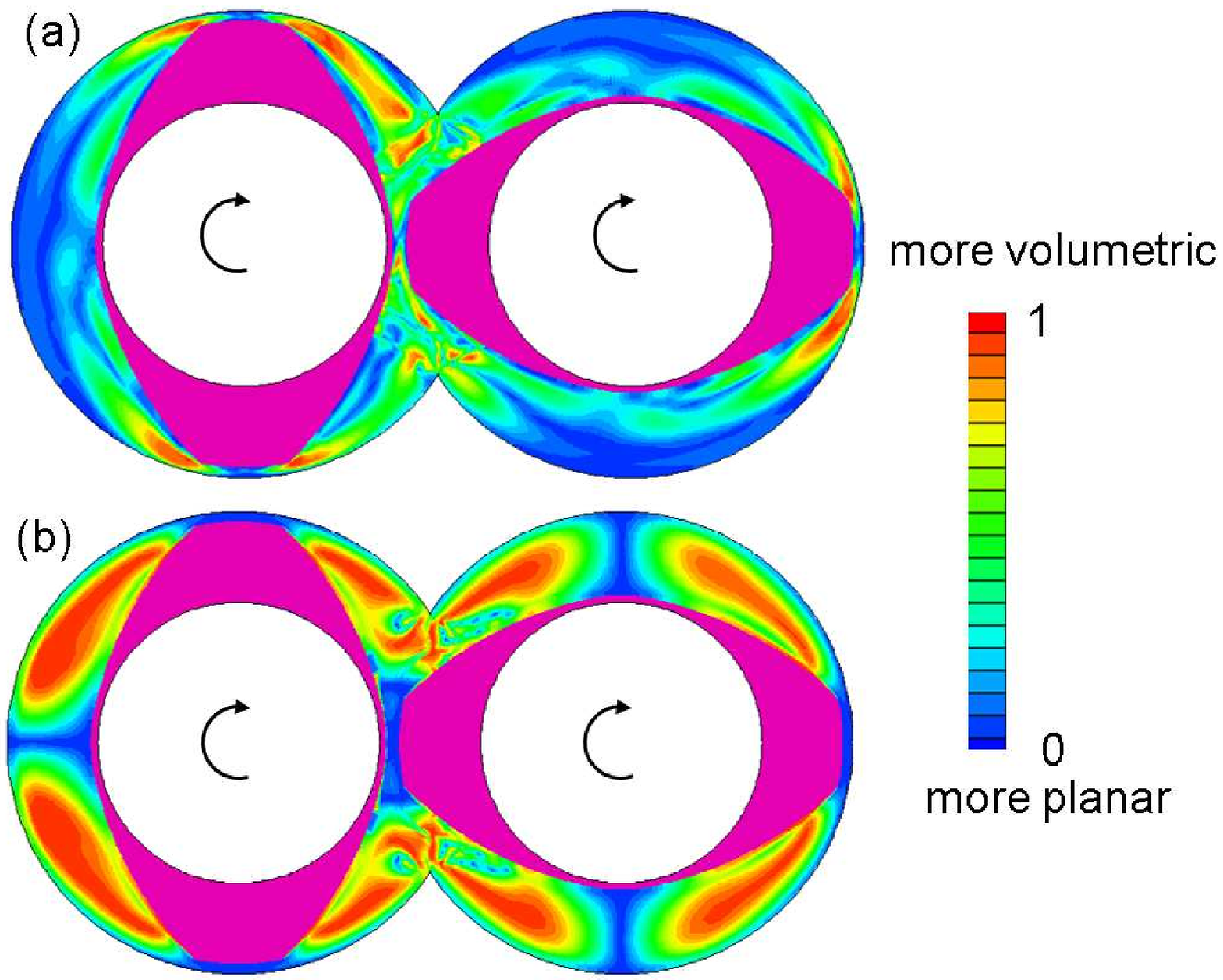}
\caption{\texttt{\figthree}
}
\label{fig3}
\end{figure}

As for the kneading disks, 
the characteristic pattern in the distribution of \(d_{v}\) is supposed
to be related to their known good mixing capability~\cite{Bravo2004Study,Lawal1995Mechanisms,Carneiro1999Flow,Funatsu20023D,Zhang2009Numerical,SarhangiFard2013Simulation}.
For understanding the relation between the distribution of \(d_{v}\) and
the mixing capability, we discuss the flow pattern in the
small strain rate region in the kneading zone.
Figure~\ref{fig4}(a) shows the velocity field at the mid-plane of the
channel at the third disk in the kneading zone.
From Fig.~\ref{fig4}(a), we see
the flow trajectories from a tip to two neighboring disks along the
screw rotation. 
Simultaneously, converging flows from two neighboring
disks are developed behind another side of screw tips.
In other words, the flow driven by the kneading disks bifurcates into forward and
backward extrusion directions on the one place and converges from
neighboring disks on the other place.
Because of these bifurcated trajectories, the fluid elements go back and
forth within the zone of consecutively arranged disks, and repeatedly
bifurcate and converge having many chances to be stretched and folded 
resulting in the high mixing capability of the kneading disks.
This property of the flow pattern is reflected by the distribution of
the biaxial/uniaxial elongational flow from the viewpoint of the strain rate mode.
Figure~\ref{fig4}(b) shows the distribution of \(d_{v}\) corresponding
to the velocity field in Fig.~\ref{fig4}(a) and clearly captures this
characteristic flow pattern structure developed in the small strain rate
region of the kneading disks.
This analysis demonstrates that the volumetric strain rate distribution
can be useful to characterize the flow pattern structure, especially in
the small strain rate regions.
\newcommand{\figfour}{Flow pattern in an internal section of a small strain rate region in the
 kneading disks:
(a) velocity field, and 
(b) \protect\(d_{v}=\protect\left|\protect\beta\protect\right|\protect\) distribution.
In (a), the thin arrows indicate the velocity field, while the big
 arrows are an eye-guide indicating the flow trends.}

\begin{figure}
 \center
\includegraphics[width=\hsize]{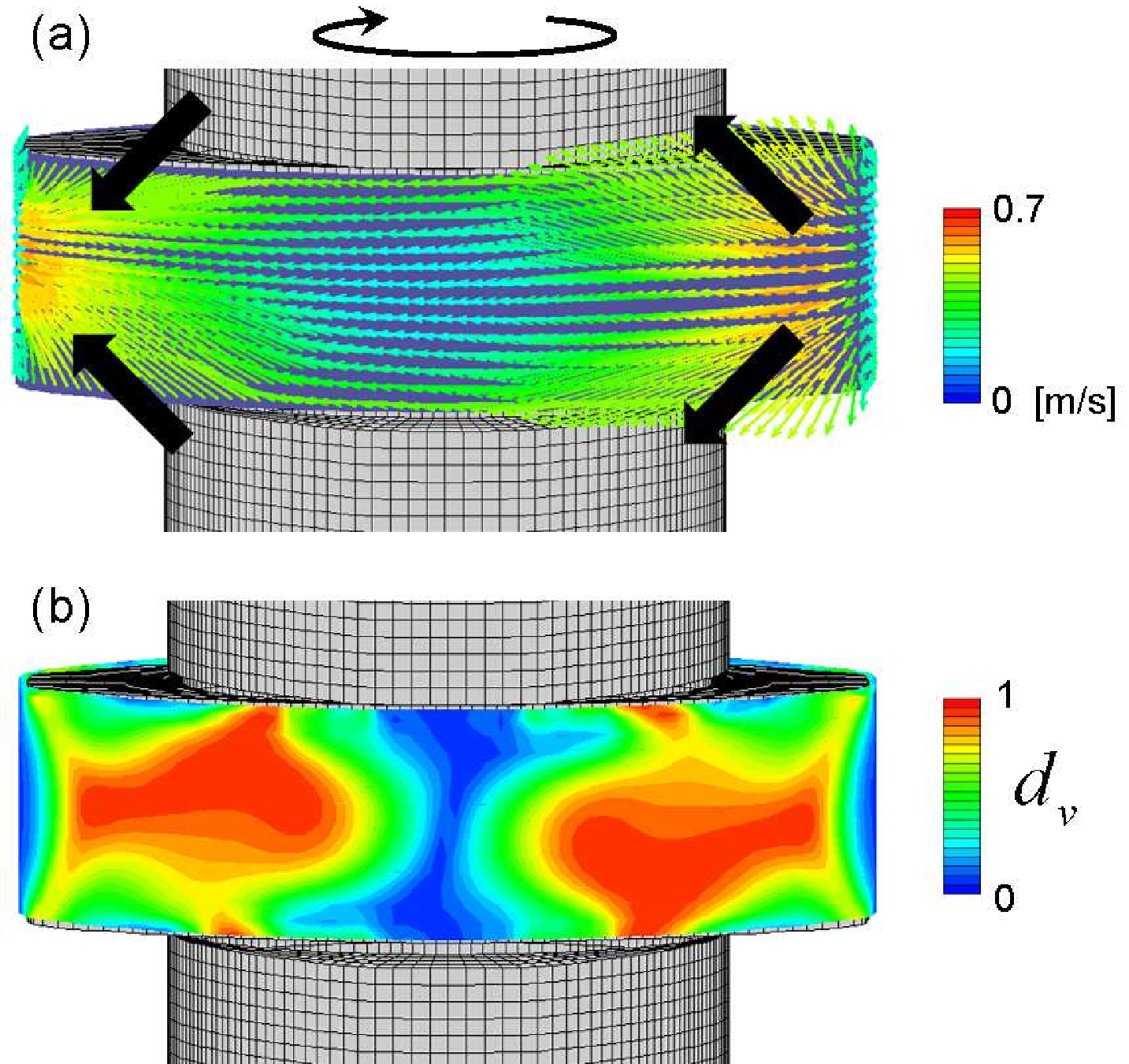}
\caption{\texttt{\figfour}
}
\label{fig4}
\end{figure}

For comparison, the distribution of the mixing index (or the
irrotationality), \(\lambda\),  of Eq.~(\ref{eq:mz})
at the cross
sections A and B in Fig.~\ref{fig1} are shown in Fig.~\ref{fig5}.
We found that both the conveying screw and the kneading disks share some
common characteristics in the \(\lambda\) distribution.
The deformation rate is almost half rotational
near the surfaces of the screws and the barrel, 
while it is almost irrotational in the small strain rate region far from
the surfaces.
Although the irrotational regions for both elements are different,
depending on their geometric shapes, the \(\lambda\) distribution does
not reflect the flow pattern shown in Fig.~\ref{fig4}(a), suggesting
that the mixing index itself hardly offers insight into the flow pattern
structure in general three-dimensional flow.

\newcommand{\figfive}{Distributions of the irrotationality
\protect\(\protect\lambda\protect\) in the melt-mixing zone.  The value
of \protect\(\protect\lambda\protect\) is zero for a pure rotation (or
zero strain rate), and unity for an irrotational one.  (a) The section
in a conveying screw at position B in Fig.~\ref{fig1}, (b) The
section in the kneading disks at position A in Fig.~\ref{fig1}.}

\begin{figure}
 \center
\includegraphics[width=\hsize]{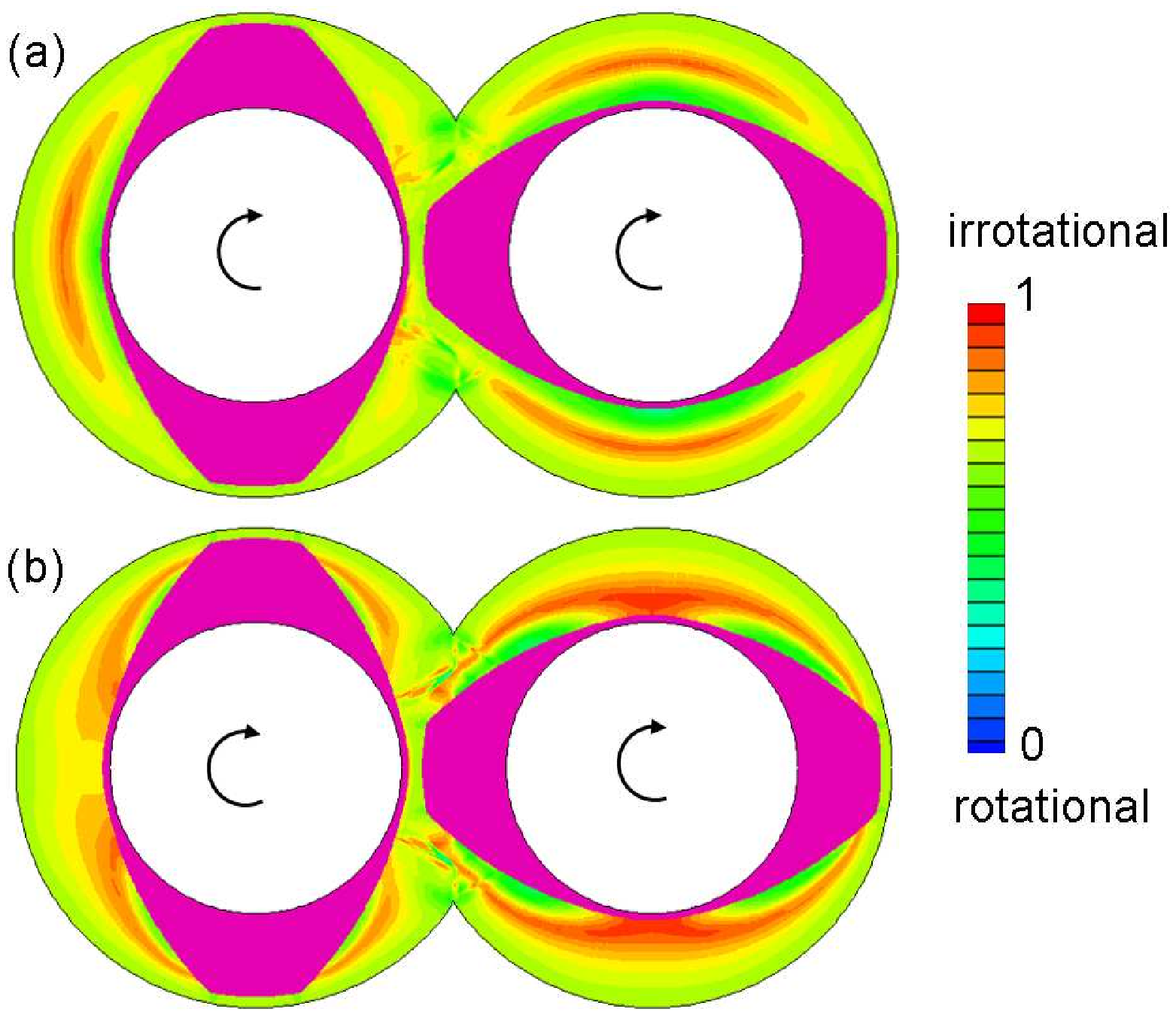}
\caption{\texttt{\figfive}
}
\label{fig5}
\end{figure}

In order to discuss the large scale characteristics of the screw
elements, \(d_{v}\) and \(\lambda\) are averaged in each section and
over one screw rotation. The axial profiles of the mean \(d_{v}\) and
\(\lambda\) are shown in Fig.~\ref{fig6}.
In proximity to the inlet and the outlet, the mean values are
supposed to be affected by the boundary conditions, and so, physically
irrelevant. 
We hence consider the axial locations from 20\;mm to 200\;mm.
The means of \(d_{v}\) and \(\lambda\) remain at an almost constant level in the
regions of the two conveying screws.
In addition, these mean quantities do not vary much at the first and
last blocks in the kneading disks because they are rather smoothly
connected to the conveying screws.
In contrast, the means of \(d_{v}\) and \(\lambda\) show large variation
along the inner three blocks of the kneading disks. 
The piecewise variations of these values are due to the piecewise
structure of the kneading disks.
In particular, the variation in the mean \(d_{v}\) within the inner
three blocks is remarkable. The mean \(d_{v}\) has a peak at the center
of each of the three blocks, which reflects the bifurcated flow trajectories back
and forth as observed in Fig.~\ref{fig4}.
The axial profile of the mean \(d_{v}\) clearly shows that the flow
pattern structure specific to the kneading disks occurs in the inner blocks and
originates from the consecutive staggered arrangement of the blocks. 
As demonstrated above, a flow pattern being effective for distributive mixing is
closely related to the distribution of the volumetric strain rate
which indicates
three-dimensional bifurcation and converging of trajectories.
The distribution of \(d_{v}\) offers a physical insight into the 
high mixing capability of the kneading disks as well as the low mixing
capability of the conveying screws.

\newcommand{\figsix}{Mean degree of volumetric strain rate and mean
irrotationality (or mixing index) as functions of axial location.  The
values of \protect\(d_{v}\protect\) and
\protect\(\protect\lambda\protect\) are averaged spatio-temporally in
sections perpendicular to the extrusion direction and over a period of
one screw rotation. The region of each element, namely the forwarding
conveying screw~(FS), the neutrally-staggered kneading disks~(N-KD), and
the backwarding conveying screw~(BS), are indicated by the left right
arrow.
}
\begin{figure}
 \center
\includegraphics[width=\hsize]{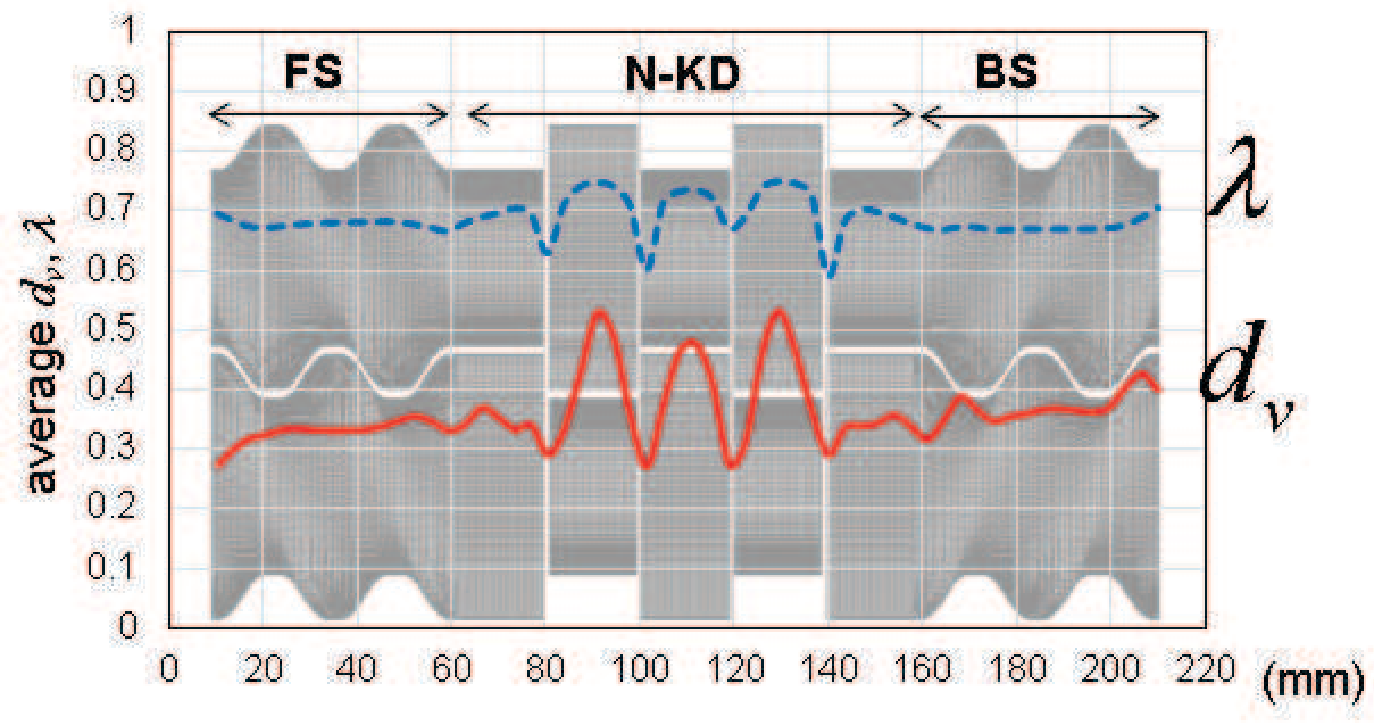}
\caption{\texttt{\figsix}
}
\label{fig6}
\end{figure}

\section{Conclusions}
We derived a scalar measure, \(\beta\), which characterizes the mode of the local
strain rate by combining the second and the third invariants of the
strain rate tensor.
Using the value of \(\beta\), the mode of the strain rate tensor 
including a uniaxial/biaxial elongational flow, a planar shear flow, and
arbitrary combinations, is defined for the general three-dimensional flows observed in fluid
mechanical processes
without solving the eigenvalue problem for the strain rate tensor.

The spatial distribution of
non-planar, or volumetric, strain rate is closely related to the flow
patterns, irrespective of the magnitude of the strain rate, and
therefore it was found to be useful to understand the relation between
the mixer geometry and the flow pattern structures. 
From the viewpoint of mixing processes, the flow pattern in the small
strain rate regions has an important role in efficient and repetitive
transport of the fluid elements to the large strain rate regions in
order to enhance the net mixing capability of the mixer. 
For understanding the effectiveness of the flow pattern especially in the small
strain rate regions, the distribution of the volumetric strain rates can
be a useful tool.
Based on the numerical simulation of a melt-mixing flow in twin-screw
extrusion, flows driven by the conveying screws and the kneading disks
have been analyzed. We found that the flow patterns specific to these elements are
clearly characterized by the distribution of the strain rate mode. 

Understanding the relation between the geometric structure of the mixing
elements and the flow pattern they drive is an important issue
in the essential evaluation of the mixing capability of the mixing devices used
in different industries.
The analysis employing the strain rate mode and its distribution is effective
for discussing the flow pattern in the mixing device and offers an insight
into the role of the small strain rate regions on the distributive mixing.
Finally, we would like to mention a limitation of the analysis only by
\(\beta\) for the mixing.
Although the distribution of the volumetric strain rates can give
information about the flow pattern structures specifying the regions where the
flow bifurcates and converges, it can just distinguish the potential of
particular flows.
For direct evaluation of the mixing, kinetic evolution of interface is
needed to discuss the area growth, interface folding, and material distribution.
Computation of the interface kinetics requires the directions of the
area segments, and therefore the mixing is a function of the orientation
of the interface relative to the flow.
This aspect of the mixing is another important problem from the
evaluation of the potential of the flow.
The combined use of our measure for the strain rate mode with 
other fluid mechanical analyses, including kinetic evolution of 
area elements is an important future direction of research into 
predicting the mixing capabilities of different mixing elements 
and in designing improved novel mixing elements.


\bigskip

\section*{Acknowledgments}
The numerical calculations have been partly carried out using the
computer facilities at the Research Institute for Information Technology
at Kyushu University.
This work has been supported by Grants-in-Aid for Scientific Research
(JSPS KAKENHI) under Grants Nos.~26400433, 24656473, and 15H04175.

\renewcommand{\refname} {Literature Cited}


%
\end{document}